\documentclass[12pt]{article} 
\usepackage[dvips]{graphicx}
\begin{document}
\title{Harris sheet solution for magnetized quantum plasmas}
\author{F. Haas\footnote{ferhaas@exatas.unisinos.br}\\
Universidade do Vale do Rio dos Sinos - UNISINOS \\
Unidade de Exatas e Tecnol\'ogicas \\
Av. Unisinos, 950\\
93022--000 S\~ao Leopoldo, RS, Brazil}
\maketitle
\begin{abstract}
We construct an infinite family of one-dimensional equilibrium
solutions for purely magnetized quantum plasmas described by the
quantum hydrodynamic model. The equilibria depends on the solution
of a third-order ordinary differential equation, which is written
in terms of two free functions. One of these free functions is
associated to the magnetic field configuration, while the
other is specified by an equation of state. The case of a Harris
sheet type magnetic field, together with an isothermal
distribution, is treated in detail. In contrast to the classical Harris sheet solution, the quantum case exhibits an oscillatory pattern for the
density.
\end{abstract}
\section{Introduction}
Quantum plasmas have attracted renewed attention in the last years, due e.g. to the relevance of quantum effects in 
ultra-small semiconductor devices \cite{Markowich}, dense plasmas \cite{Jung} and very intense laser plasmas \cite{Kremp}. The most recent developments in collective effects in quantum plasmas comprises wave propagation in dusty plasmas \cite{Stenflo}-\cite{Misra}, soliton and vortex solutions \cite{Shukla1, Yang}, shielding effects \cite{Shukla3, Ali2}, modulational instabilities \cite{Marklund} and spin effects \cite{Marklund2}. Most of these works have been made using the hydrodynamic model for quantum plasmas \cite{Haas1}-\cite{Manfredi1}, in contrast to more traditional approaches based on kinetic descriptions \cite{Pines}. Microscopic descriptions like coupled Schr\"odinger equations or
Wigner function approaches are more expensive, both numerically and
analytically, specially if magnetic fields are allowed. For a general review on the available quantum plasma models, see \cite{Manfredi2}. 

The electrostatic fluid model for quantum plasmas have been recently extended to incorporate magnetic fields \cite{Haas4}. The new quantum hydrodynamic model was derived taking the first two
moments of the electromagnetic Wigner equation, which is the quantum counterpart of the corresponding 
Vlasov equation, and assuming a closure condition $p = p(n)$. In
other words, the procedure is formally the same as for classical
plasma fluid descriptions, while now the starting point is the
Wigner-Maxwell and not the Vlasov-Maxwell system. The electromagnetic quantum fluid model has been already used for 
the analysis of shear Alfv\'en modes in ultra-cold quantum magnetoplasmas \cite{Shukla2}, the description of drift modes in nonuniform quantum magnetoplasmas \cite{Shukla4} and of shear electromagnetic waves in electron-positron plasmas \cite{Shukla8}. Instead of the discussion of wave propagation in quantum plasmas, the aim of this letter is the analysis of some simple quantum magnetostatic equilibria resembling the well known Harris profile for classical plasma \cite{Harris}. 

\section{Quantum magnetoplasma equilibria}

For a one-component quantum plasma, the
electromagnetic quantum fluid equations reads \cite{Haas4}
\begin{eqnarray}
\label{e1}
\frac{\partial n}{\partial t} + \nabla\cdot(n {\bf u}) &=& 0 \,,\\
\frac{\partial{\bf u}}{\partial t} + {\bf
u}\cdot\nabla{\bf u} &=& - \frac{1}{mn}\nabla\,p - \frac{e}{m}({\bf
E} + {\bf u}\times{\bf B})  \nonumber \\ \label{e2} &+&
\frac{\hbar^{2}}{2m^2}\nabla\left(\frac{\nabla^{2}\sqrt{n}}{\sqrt{n}}\right)
\,.
\end{eqnarray}
All the symbols in eqs. (\ref{e1}-\ref{e2}) have their conventional meaning and the system is supplemented by Maxwell equations. Only electrons are considered, the ions being described by a convenient immobile background. Notice
the extra dispersive term, proportional to $\hbar^2$, at the
moment transport equation. This Bohm potential term has profound consequences on
the structure of the equilibrium solutions, as we shall see in the
following.

Specifically, consider a purely magnetic one-dimensional class of
time-independent solutions characterized by zero electric field
and
\begin{eqnarray}
{\bf B} &=& B_{y}(x)\hat{y} + B_{z}(x)\hat{z} \,,  \nonumber \\
\label{e3} n &=&  n(x) \,, \\ {\bf u} &=&
u_{y}(x)\hat{y} + u_{z}(x)\hat{z}  \,, \nonumber \\
p &=& p(n) \,. \nonumber
\end{eqnarray}
The magnetic field can be given in terms of a vector potential
${\bf A} = A_{y}(x)\hat{y} + A_{z}(x)\hat{z}$, so that $B_y = -
dA_{z}/dx$ and $B_z = dA_{y}/dx$. Notice that the fluid model is suitable for the search for 
static quantum equilibria since the kinetic (Wigner) equation is not satisfied by arbitrary functions 
of the invariants of motion as for Vlasov plasmas. Therefore we are not allowed to use Jeans theorem for 
the construction of equilibria. 

Neutrality is enforced by an appropriate immobile ionic background described by an ionic density $n_{i}(x)$. Therefore, Poisson's equation can be ignored. Now inserting the form
(\ref{e3}) into Amp\`ere's law and the quantum fluid equations
gives
\begin{eqnarray}
\label{e4}
\frac{d^{2}A_y}{dx^2} &=&  e \mu_{0}n u_y \,, \\
\label{e5}
\frac{d^{2}A_z}{dx^2} &=&  e \mu_{0}n u_z \,, \\
\frac{dp}{dx} &=& - e n (u_y \frac{dA_y}{dx} + u_z
\frac{dA_z}{dx}) \nonumber \\ \label{e6} &+&
\frac{\hbar^{2}n}{2m}\frac{d}{dx}\left(\frac{d^{2}\sqrt{n}/dx^2}{\sqrt{n}}\right)
\,.
\end{eqnarray}

As in the classical situation \cite{Attico}, it is useful to
restrict to the cases where the magnetic field is indirectly
defined through a pseudo-potential $V = V(A_{y},A_{z})$ for which
\begin{eqnarray}
\label{e7}  n u_y = - \frac{1}{e\mu_0}\frac{\partial V}{\partial A_y} \,,\\
\label{e8} \quad  n u_z = - \frac{1}{e\mu_0}\frac{\partial
V}{\partial A_z} \,,
\end{eqnarray}
so that (\ref{e4}-\ref{e5}) is transformed into a two-dimensional
autonomous Hamiltonian system,
\begin{eqnarray}
\label{e9} \frac{d^{2}A_y}{dx^2} &=& - \frac{\partial V}{\partial
A_y} \,, \\ \label{e10} \frac{d^{2}A_z}{dx^2} &=& - \frac{\partial
V}{\partial A_z} \,.
\end{eqnarray}
In this system, $x$ plays the r\^ole of time, while the components
of the vector potential play the r\^ole of spatial coordinates.
After specifying the pseudo-potential $V$ and solving eqs.
(\ref{e9}-\ref{e10}), we regain the magnetic field using ${\bf B}
= \nabla\times{\bf A}$. The current then follows from eqs. 
(\ref{e7}-\ref{e8}).

The choice expressed at eqs. (\ref{e7}-\ref{e8}) imposes a
restriction on the classes of equilibria, since not all density
and velocity fields can be cast in this potential form. However,
introducing the pseudo-potential $V$ has at least two advantages.
First, we can learn from Hamiltonian dynamics how to design
specific pseudo-potentials $V$ in order to obtain special classes of
magnetic fields. For instance, periodic magnetic fields can be
easily obtained from well known potentials associated to periodic
solutions. Second, the formalism becomes more compact in terms of
the function $V$.

In terms of $V$, the balance eq. (\ref{e6}) reads
\begin{equation}
\label{eqq11} \frac{d}{dx}\left(p - \frac{V}{\mu_0}\right) =
\frac{\hbar^{2}n}{2m}\frac{d}{dx}\left(\frac{d^{2}\sqrt{n}/dx^2}{\sqrt{n}}\right)
\,.
\end{equation}
It can be shown that, apart from an irrelevant numerical
constant, the pseudo-potential $V$ is directly related to magnetic pressure, $V = - |{\bf B}|^{2}/2$,
showing that the left-hand side of eq. (\ref{eqq11}) refers to the usual (classical) pressure
balance equation. The right-hand side, however, has a pure
quantum nature. Not only there must be a balance between kinetic
and magnetic pressures, since the quantum pressure arising from
the Bohm potential term has to be taken into account. This quantum
pressure manifests e.g. in the dispersion of wave-packets in standard quantum mechanics. In plasmas, the quantum
pressure is responsible for subtle effects like in the case of the
quantum two-stream instability, where the instability is magnified
for small wave-numbers and suppressed for large wave-numbers
\cite{Haas1, Manfredi1}.

In the quantum case where $\hbar \neq 0$, eq. (\ref{eqq11}) is
a third-order ODE for the density. It is useful to express this
equation in terms of a variable $a = \sqrt{n}$. Taking into
account the equation of state $p = p(n)$ and defining a new
function $\tilde{V}(x) = V(A_{y}(x),A_{z}(x))$, we get
\begin{equation}
\label{eqq12} a a''' - a' a'' + f(a) a' + g(x) = 0 \,,
\end{equation}
where the prime denotes differentiation with respect to $x$ and we
have introduced the quantities
\begin{eqnarray}
\label{eqq13}
f(a) &=& - \frac{4ma}{\hbar^2} \frac{dp}{dn}(n = a^2) \,,\\
\label{eqq14} g(x) &=&
\frac{2m}{\mu_{0}\hbar^2}\frac{d\tilde{V}}{dx} \,.
\end{eqnarray}
The strategy to derive the solutions is now clear. Choosing a
pseudo-potential $V(A_{y},A_{z})$ and then solving the Hamiltonian system
(\ref{e9}-\ref{e10}) for the vector potential, we determine
simultaneously the magnetic field and $\tilde{V}$. Quantum effects manifests in the
equation for the density, eq. (\ref{eqq12}), which also
deserves the function of state $p = p(n)$.

Another legitimate interpretation of the balance equation
(\ref{eqq12}) is to first specify the particle density $n$ and the
magnetic pressure $-V/\mu_0$ and then solving for the kinetic
pressure. This would give an equation of state with a quantum
correction. However, in most applications, one supposes a certain
equation of state and then proceeds to the calculation of the
density and velocity fields. This will be our preferred approach
in what follows. In the next section, we consider in detail the case of Harris
sheet magnetic fields.

\section{Quantum Harris sheet}
Exactly as for the classical Harris solution, suppose a isothermal plasma, $p = n\kappa_{B}T$, and a pseudo-potential function
\begin{equation}
\label{eq11}
V = \frac{B_{\infty}^2}{2}\exp(\frac{2A_z}{B_{\infty}L}) \,,
\end{equation}
where $L$ is a characteristic length and $B_\infty$ is a
(constant) magnetic field reference value. The Hamiltonian system
(\ref{e9}-\ref{e10}) is then
\begin{equation}
\label{eq12}
\frac{d^{2}A_y}{dx^2} = 0 \,, \quad \frac{d^{2}A_z}{dx^2} = - \frac{B_\infty}{L}\exp(\frac{2A_z}{B_{\infty}L}) \,.
\end{equation}
If we further take the boundary conditions $A_{z}(x = 0) = (dA_{z}/dx)(x = 0) = 0$, we easily solve (\ref{eq12}) to get
\begin{equation}
A_y = A_{y0} + B_{0} x \,, \quad A_z = - B_{\infty}L \,\ln\cosh(x/L) \,,
\end{equation}
where $A_{y0}$ and $B_0$ are integration constants. The magnetic field following from this vector potential characterizes the well-known Harris sheet solution,
\begin{equation}
B_y = B_{\infty}\tanh(x/L) \,, \quad B_z = B_{0} \,,
\end{equation}
also allowing for a superimposed homogeneous magnetic field.

In addition, the velocity field follows from (\ref{e4}-\ref{e5}),
\begin{equation}
\label{u}
u_y = 0 \,, \quad u_z = \frac{B_{\infty}}{e\mu_0 nL}\sec\!{\rm
h}^{2}(\frac{x}{L}) \,.
\end{equation}
Notice that any departure from the classical density solution would imply further changes in the velocity field.

To derive the density we have to solve the third-order ODE eq.
(\ref{eqq12}), constructed in terms of the functions $f(a)$ and
$g(x)$ at (\ref{eqq13}-\ref{eqq14}). Using the isothermal equation of
state, the form (\ref{eq11}) for the pseudo-potential
$V$ and the Harris sheet solution, we get
\begin{eqnarray}
f(a) &=& - \frac{4m\kappa_{B}T}{\hbar^2}\,a \,, \\
g(x) &=& -
\frac{mB_{\infty}^2}{\mu_{0}\hbar^{2}L}\sec\!{\rm
h}^{2}(\frac{x}{L})\tanh(\frac{x}{L}) \,.
\end{eqnarray}

Adopting the dimensionless variables
\begin{equation}
\label{resc}
\alpha = a/\sqrt{n_0} \,, \quad X = x/L \,,
\end{equation}
where $n_0$ is some ambient density such that 
\begin{equation}
\label{eq14}
n_0 \kappa_B T = \frac{B_{\infty}^2}{4\mu_0} \,,
\end{equation}
eq. (\ref{e8}) is finally expressed as
\begin{equation}
\label{e15} \alpha\frac{d^{3}\alpha}{dX^3} -
\frac{d\alpha}{dX}\frac{d^{2}\alpha}{dX^2} -
\frac{\alpha}{H^2}\,\frac{d\alpha}{dX} =
\frac{1}{H^2} \, \sec\!{\rm
h}^{2}X \tanh\,X \,,
\end{equation}
in terms of a new dimensionless parameter
\begin{equation}
\label{e16}
H = \frac{\hbar}{mV_{a}L} \,,
\end{equation}
where $V_a = B_{\infty}/(\mu_{0}mn_{0})^{1/2}$ is the Alfv\'en velocity.

The parameter $H$ is a measure of the relevance of the quantum
effects. It is essentially the ratio of the scaled Planck constant
$\hbar$ to the action of a particle of mass $m$ travelling with
the Alfv\'en velocity and confined in a length $L$ related to 
the thickness of the sheet. The larger the ambient density $n_0$
and the smaller the characteristic length $L$ or the
characteristic magnetic field $B_\infty$, the larger are the
quantum effects.

In order to understand the r\^ole of the quantum terms, we may
investigate (\ref{e15}) with 
\begin{equation}
\label{e17} \alpha(X = 0) = 1 \,, \quad \frac{d\alpha}{dX}(X =0) = 0 \,,
\quad \frac{d^{2}\alpha}{dX^2}(X = 0) = - 1 \,,
\end{equation}
which reproduces the boundary conditions for the classical Harris
sheet, when $\alpha = \sec\!{\rm h}X$. With the choice (\ref{e17}),
eq. (\ref{e15}) integrated once gives
\begin{equation}
\label{x} \alpha\frac{d^{2}\alpha}{dX^2}- \left(\frac{d\alpha}{dX}\right)^2 + 1 =
\frac{1}{2H^2}\,\left(\alpha^2 - \sec\!{\rm h}^{2}X\right) \,.
\end{equation}
In the ultra-quantum limit $H \rightarrow \infty$, the left-hand side of (\ref{x}) vanishes. In this situation and using the prescribed boundary conditions, the solution is
\begin{equation}
\alpha = \cos X \,.
\end{equation}
This imply a qualitative change (from localized to oscillatory) on
the solution due to quantum effects. In order to further
investigate this tendency, we show the numerical solution for
(\ref{x}) with the appropriate boundary conditions for a few
values of $H$. This is shown in the figs. 1 and 2, where
increasingly oscillatory solutions are shown, according to $H = 1$ or $H = 5$ respectively. On the opposite
case, (\ref{x}) shows that when $H \rightarrow 0$ we regain the classical Harris
solution, $\alpha = \sec\!{\rm h}X$.

\begin{figure*}
\begin{center}
\includegraphics{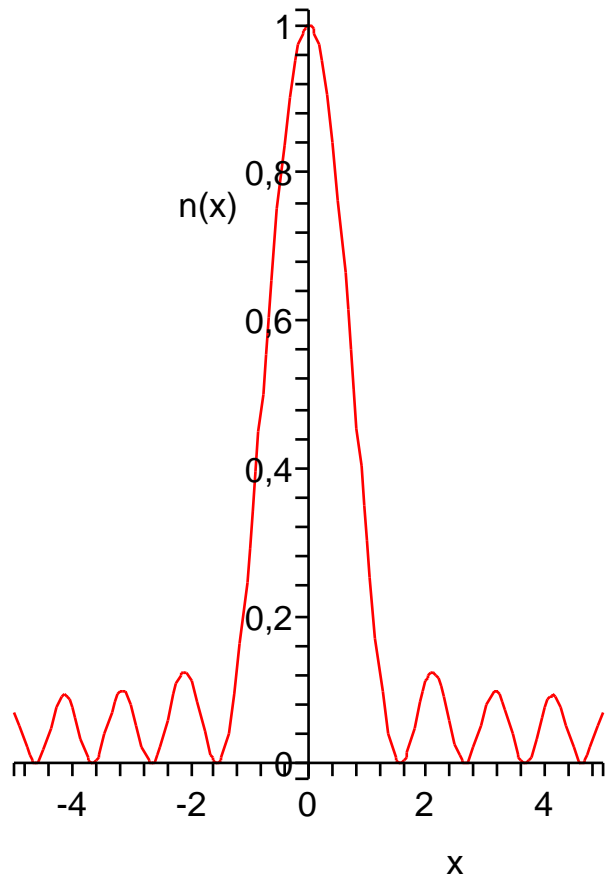}
\caption{Density oscillations for $H = 1$. Parameters: $n_0 = L = 1$.}
\label{f.1}
\end{center}
\end{figure*}

\begin{figure*}
\begin{center}
\includegraphics{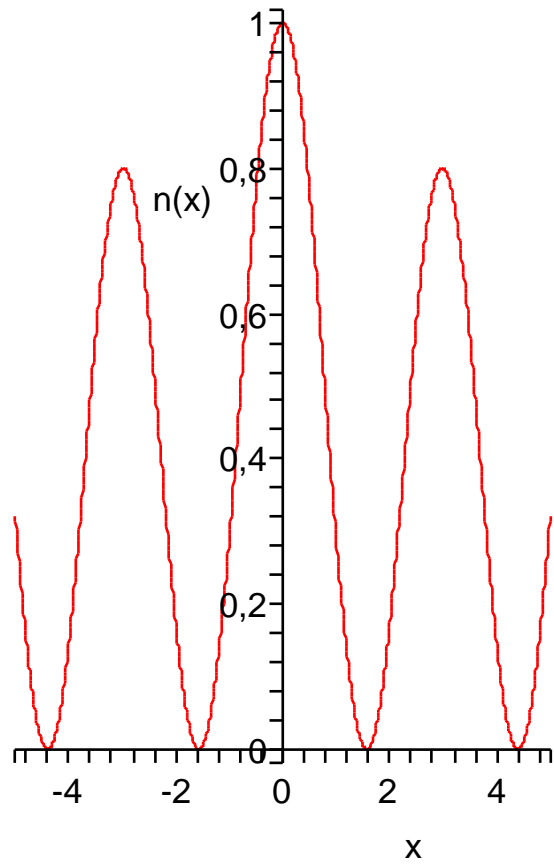}
\caption{Density oscillations for $H = 5$. Parameters: $n_0 = L = 1$.}
\label{f.2}
\end{center}
\end{figure*}

Another interesting possibility is an equation of state for an ultra-cold Fermi gas,
\begin{equation}
p = \frac{2\kappa_{B}T_F}{5n_{0}^{2/3}}\,n^{5/3} \,,
\end{equation}
where $T_F$ is the Fermi temperature and $n_0$ is the ambient density. Proceeding exactly as before and assuming 
\begin{equation}
n_{0}\kappa_{B}T_F = \frac{3B_{\infty}^2}{8\mu_0} \,,
\end{equation}
we obtain 
\begin{equation}
\alpha\frac{d^{3}\alpha}{dX^3} -
\frac{d\alpha}{dX}\frac{d^{2}\alpha}{dX^2} -
\frac{\alpha^{7/3}}{H^2}\,\frac{d\alpha}{dX} =
\frac{1}{H^2} \, \sec\!{\rm
h}^{2}X \tanh\,X \,,
\end{equation}
where $\alpha$ and $X$ are defined in (\ref{resc})and $H$ in (\ref{e16}). Similar oscillatory behaviour is also found for nonzero $H$ and suitable boundary conditions. 

\section{Summary}
Equation (\ref{eqq12}) describes a whole class of quantum equilibria for magnetoplasmas. 
The particle density compatible with a $tanh$ magnetic
field shows an increasingly oscillatory pattern, in comparison to the
classical system associated to a localized $sech^2$ solution. Other classes of equilibria can be built for 
different choices of pseudo-potentials $V = V(A_{y},A_{z})$ and equations of state $p = p(n)$.
The ideas in the present formulation may be a starting point for  magnetic equilibria relevant for dense astrophysical objects like white dwarfs. 

\noindent
{\bf Acknowledgments}

\noindent
We thanks the Brazilian agency Conselho Nacional de
Desenvolvimento Cien\-t\'{\i}\-fi\-co e Tec\-no\-l\'o\-gi\-co (CNPq) for financial support. We also thanks Prof. VINOD KRISHAN for useful comments.

\end{document}